\documentclass[runningheads]{llncs}
\usepackage[T1]{fontenc}
\usepackage{graphicx}
\begin{document}
%
\title{Shared SAT Solvers and SAT Memory\\in Distributed Business Applications}
%
%
\author{Sergejs Kozlovičs
}
%
%
\institute{Institute of Mathematics and Computer Science, University of Latvia\\
Raiņa bulv. 29, Riga, Latvia, LV-1459\\
\email{sergejs.kozlovics@lumii.lv}}

\maketitle              
\begin{abstract}
We propose a software architecture where SAT solvers act as a shared network resource for distributed business applications. There can be multiple parallel SAT solvers running either on dedicated hardware (a multi-processor system or a system with a specific GPU) or in the cloud. In order to avoid complex message passing between network nodes, we introduce a novel concept of the shared SAT memory, which can be accessed (in the read/write mode) from multiple different SAT solvers and modules implementing the business logic. As a result, our architecture allows for the easy generation, diversification, and solving of SAT instances from existing high-level programming languages without the need to think about the network. We demonstrate our architecture on the use case of transforming the integer factorization problem to SAT.

\keywords{SAT \and distributed applications \and software architecture \and integer factorization}
\end{abstract}
\sloppy
\section{Introduction}

The Boolean satisfiability problem (SAT) has many practical applications such as circuit design, model generation and verification, planning, software package management, program analysis, and other constraint satisfaction problems \cite{boulanger_formal_2014,rintanen_planning_2012,eggersgluss_high_2012,bib_FznTiny}.
Since SAT is an NP-complete problem, a polynomial reduction exists for any other problem from the NP class, making SAT a "silver bullet" for the NP class\footnote{Though,, finding the most direct low-degree polynomial reduction to SAT can be a challenge.}. Even if some NP problem is not known to be NP-complete (such as integer factorization or graph isomorphism), reducing it to SAT can be a reasonable temporary measure until a specific efficient algorithm is found (if it exists).

While no polynomial algorithm is known for SAT, and we do not know whether it exists (the P=NP problem), numerous techniques have been proposed to solve SAT efficiently. They include heuristics, conflict-driven clause learning, backjumping, random restarts, message passing, and machine learning \cite{malik_boolean_2009,biere_handbook_2021,bib_sat_chaff}. As a result, state-of-the-art SAT solvers are able to find variable assignments for SAT instances with tens of thousands of variables and clauses, meaning that many practical SAT applications are now tractable. 

While some solvers are optimized to run on a single-core CPU (e.g., Glusoce and GSAT), others take advantage specific of hardware, e.g., HordeSAT utilizes multiple CPU cores \cite{audemard_glucose_2018,folino_parallel_2001,balyo_hordesat_2015}. At IMCS\footnote{Institute of Mathematics and Computer Science, University of Latvia}, we have developed our own solver QuerySAT, which uses graph neural networks internally and requires a GPU\footnote{such as Nvidia T4 16GB GPU (used in our experiments)} for better performance \cite{ozolins_goal-aware_2022}.

Solvers that use branching (e.g., MapleSAT\footnote{MapleSAT has multiple configurations. It is based on MiniSAT \cite{bib_minisat};\\ see \url{https://sites.google.com/a/gsd.uwaterloo.ca/maplesat/maplesat}.} and Lingeling\footnote{\url{https://github.com/arminbiere/lingeling}}) can be used together with parallel solvers that can diversify a single SAT instance into multiple subtasks to be executed in parallel. Besides, multiple different solvers can be launched in parallel on the same SAT instance, hoping that one of them will find the solution faster. Furthermore, we can combine solvers, resulting in hybrid ones, e.g., we can replace the query mechanism in our QuerySAT with a third-party solver.

Since SAT solvers are computationally intense and often require specific hardware, we consider the task of making SAT solvers available as a shared resource deployed either to the dedicated on-premise hardware on to the cloud servers. In this paper, we cover the following tasks:
\begin{enumerate}
\item Integrating SAT solvers (as a shared resource) into distributed business software.
\item Optimizing the communication between multiple parallel SAT solvers (e.g., for diversification and learned clause exchange).
\end{enumerate}

Our idea is to introduce the shared SAT memory, which can be accessed from all involved modules (SAT solvers and business software components).

The next section provides the terminology and introduces the web kernel concept-- an OS kernel analog for distributed applications. The web kernel will act as the communication broker between SAT solvers, business logic units, and SAT memory. We continue by providing essential implementation details. In particular, we show how to implement SAT memory and manage SAT solvers on multiple shared hardware units. In Section \ref{sec:usage}, we provide a usage scenario based on integer factorization.
Finally, we discuss related work and sketch further research directions.



\section{Definitions}
A {\bf distributed application} is a program that runs on more than one autonomous computer that communicate over a network. Usually, it consists of two separate programs: the back-end (server-side software) and the front-end (client-side) software. However, multiple server and client nodes are also possible.

The {\bf web computer} is an abstraction that simplifies the development of distributed applications by providing the illusion of a single target computer~\cite{kozlovics_web_2019,kozlovics_webappos_2020}. The web computer factors out the network and multi-user/multi-process management. As a result, distributed applications can be created like traditional desktop applications, i.e., by focusing on just one program, one target computer, and one user.

The web computer consists of the following main parts:
\begin{itemize}
\item {\bf data memory} (or web memory), which is shared among network nodes (e.g., the client and the server); since it is constantly being automatically and transparently synchronized, each network node can access web memory as if it was directly attached;
\item the {\bf instruction memory} (or code space), where executable/interpretable code is placed; the particular code is delivered only the nodes being able to execute/interpret it (e.g., Java and Python code is installed at the server-side and executed there, while JavaScript code can be delivered and executed at the client browser as well as at the server-side by means of a JavaScript engine such as node.js);
\item {\bf web processors}, which are software modules that can execute certain types of instructions from the code space;
\item {\bf web I/O devices}, which can be either physical devices (e.g., printers) directly attached to particular physical nodes and available via specific APIs, or virtual devices implemented as software running either on a single node or in a cluster (e.g., databases or file systems).
\end{itemize}

Notice that due to security considerations, the web computer separates instruction memory from data memory. Thus, it corresponds to the Harvard architecture (as opposed to von Neumann architecture). That protects server-side code from being able to execute code from the shared data memory, which the client can modify (and we should not trust the client)~\cite{bib_break_web_software}.

Although the web computer still requires multiple physical network nodes to operate, web applications do not access them directly but via the intermediate layer, making the single computer illusion possible. Since this intermediate layer acts as an operating system analog, its reference implementation is called {\em webAppOS} (available at \url{http://webappos.org}).

\begin{figure}
    \centering
    \includegraphics[width=\textwidth]{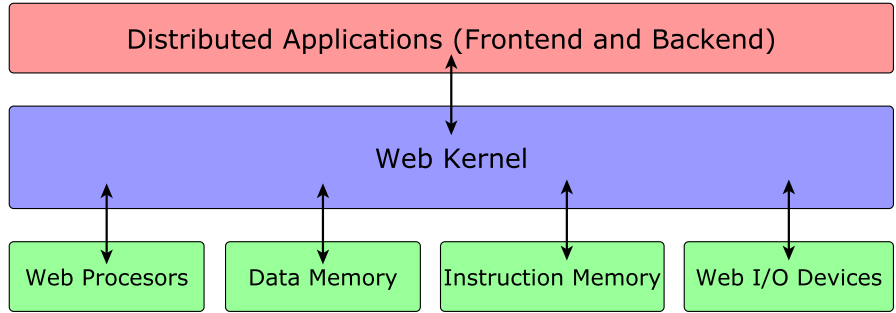}
    \caption{webAppOS web kernel as an abstraction that factors out the network}
    \label{fig:webkernel}
\end{figure}
The {\bf web kernel} is the abstraction layer used by webAppOS applications to access web memory, to invoke code from the instruction memory (the code will be executed by some web processor), and to access web I/O devices (Figure~\ref{fig:webkernel}). The web kernel factors out the network communication and provides the illusion that all web computer components are located at the same network node as the calling application code. Since code invocations may require network communication, they are called {\bf web calls}. Web calls are enqueued and forwarded to the corresponding web processors by the web kernel. All web calls are asynchronous (however, the async/await capabilities of modern programming languages can be used as a syntax sugar to simplify asynchronous code).

In fact, web memory is a graph-like structure for storing objects, their states (attributes), and the links between these objects. Besides, web memory also stores the meta-level information (object classes, inheritance relations, etc.). Thus, web memory resembles the OOP memory used by the Java virtual machine. In web memory, multiple inheritance is supported.

The code space, in its turn, stores executable code as {\bf web methods}, which are either functions or methods (depending on a programming language). Each web method is identified by an implementation-agnostic fully-qualified name (e.g., \verb"ClassName.methodName"). Thus, a web call is specified by a fully-qualified method name, the web memory slot, the web object reference in that slot (a \verb"this" or \verb"self" analog), and a JSON object as an argument. Each web call must also return a JSON object (at least an empty JSON \verb"{}", if no result is expected). Errors (including network errors) are returned in the \verb"error" field.

Different methods of the same class can be written in different programming languages (we call that approach {\bf per-method granularity}). Obviously, there must be web processors (among all nodes) that are able to execute different types of code used in the given webAppOS application.

Since methods can be executed by different web processors (which can be located on different network nodes), it is not possible to implement static linking. The process or substituting a particular web call with the corresponding method implementation is called {\bf web linking}. Due to dynamic and distributed nature of web linking, we use the
{\bf duck typing} mechanism to define the {\em implements} relationship between web objects and interfaces.\footnote{If interface $I$ is a set of web methods $m_1, m_2, \dots,$ and the object $o$ is included in web memory classes $C_1, C_2, \dots$, then $o$ implements $I$, iff $\forall m_i\exists C_j:$ 
ClassName($C_j$).methodName($m_i$) $\in$ Code Space.}

Each webAppOS application can be used by multiple users, and each user can have multiple concurrent sessions with the same application. Each such application instance is called a {\bf web process}; it has a dedicated web memory slot identified by {\bf webPid} (in essence, UUID).

Web I/O devices are implemented as code libraries that define device-specific classes and provide implementations for the device-specific methods. For real hardware, these methods access the device via native drivers or OS calls; for virtual devices, they provide software implementation. 
Since web I/O devices are shared among all web processes, there is a dedicated web memory slot called the {\bf root web memory}, where web I/O devices (with their states) are stored. Thus, device-specific methods can be invoked via the web kernel in the same way as invoking ordinary web calls. 
Internally, however, device-specific web calls are implemented like software interrupts --- they take precedence over ordinary web calls in web processors. This is done intentionally since we may need to interrupt the previous command sent to the device or to release the device for other users as soon as possible.\footnote{This implies that web I/O devices must be implemented in a thread-safe way.}

The web kernel is already available as part of webAppOS. In our architecture, it will act as the communication broker between SAT solvers, business logic units, and the SAT memory, all of which can be located at different network nodes.

\section{Integrating SAT Solvers}
Each SAT solver is integrated as a virtual web I/O device (i.e., implemented as a software unit). In the root web memory, different types of SAT solvers are represented as subclasses of the \verb"SatSolver" superclass, while each particular solver instance is represented as an object of the corresponding subclass. Typically, the number of solver instances will correspond to the number of free CPU cores (i.e., not occupied by web processors); however, the solver type and the required hardware (e.g., GPU) can also affect the number of solvers at each node. In any case, all available solvers among all nodes can be easily detected by traversing the root web memory.

\subsection{Per-Solver Concurrency}
\label{sec:solver_api}

The SAT instance can be a standalone SAT problem, a subtask of a larger problem, or its diversified variant. Since solving SAT is a computationally intensive task, we intentionally limit each solver to processing only one SAT instance at a time. However, since the solver can be invoked by different users concurrently, it has to be implemented in a thread-safe way, returning an \verb"error" attribute when necessary.

Each solver has to provide the following web methods (=web I/O device interrupts):
\begin{itemize}
\item \verb"solve({satMemoryUrl, timeout, diversification})" --- starts the solver on a SAT instance stored in the given SAT memory using the given diversification guidelines (the SAT memory is discussed in the next section, while diversification is discussed in the following subsection).

The solver gets the clauses from the SAT memory, diversifies itself, solves the SAT instance, and returns the \verb"result" attribute equal to \verb"SAT", \verb"UNSAT", or \verb"UNKNOWN".
The latter value is returned when neither the satisfying variables assignment, nor the contradiction could be devised (e.g., for heuristics-based or randomized solvers). The \verb"UNKNOWN" value is also returned when the solver has been interrupted.
In case of \verb"SAT", the variable assignments are returned in the \verb"model" attribute (a JSON array).

Suppose the solver is already solving another SAT instance. In that case, the \verb"solve" method must return the \verb"BUSY" flag in the \verb"error" attribute or wait at most \verb"timeout" seconds for the solver to become available.

\item \verb"pause({})" --- requests the solver to interrupt the search as soon as possible with the ability to resume the search (e.g., the call stack has to be maintained);
\item \verb"resume({})" --- resumes the previously interrupted search;
\item \verb"cancel({})" --- interrupts the search for good for the SAT instance currently being processed. As a result, the currently running \verb"solve" web call must return \verb"UNKNOWN".
\end{itemize}

We assume that in case of any error (e.g., when a non-paused solver is requested to \verb"resume", or when the memory/CPU limits have been exceeded), the \verb"error" attribute is returned. Besides, since the web calls above are web I/O device interrupts, \verb"webPid" of the caller web process is always implicitly passed. That allows the solver to verify, for example, whether the \verb"pause" web call originated from the same web process that had invoked \verb"solve",~--- a helpful security precaution.

\subsection{Diversification}

The following diversification settings can be passed to the \verb"solve" web call. We borrowed them from the HordeSAT portfolio-based solver and transformed them to the JSON syntax to comply with the requirements of the web kernel \cite{balyo_hordesat_2015}. Each solver type can implement diversification differently.

\begin{itemize}
    \item \verb"rank" --- the index of this particular solver among all solvers working on the current SAT problem; as a trivial example, \verb"rank" could be used to initialize the random number generator seed;
    \item \verb"size" --- the number of solvers working on the current SAT problem;
    \item \verb"phases" --- an optional JSON object with attributes named \verb"x"$i$, e.g., \verb"x5", $1\le i\le$\#(SAT variables); for variable $x_i$, \verb"phases[x"$i$\verb"]" denotes the Boolean value to try first during the search (the variable "phase").
\end{itemize}

Notice that although each solver in the root web memory is treated as a single instance, it can rely on multiple parallel solvers (e.g., multiple GPU subprograms) inside. Such "internals solvers" are not visible to the web kernel and cannot be diversified in the way described above.

\subsection{Joining Parallel Solvers}
A solver can divide the task into subtasks and forward them to multiple parallel solvers via the \verb"parallelize", a built-in web call provided by the web kernel. The \verb"parallelize" web call takes a list $L$ of child web calls to be invoked in parallel, when possible. Technically, \verb"parallilize" stores a counter in web memory and appends the \verb"join" method (also built-in) invocation to each web call from $L$. The \verb"join" method increments the counter and stores the return value of the child web call. When the counter reaches the length of $L$, \verb"parallelize" returns the list of child return values, which can be processed by the parent solver.

\section{Implementing the SAT Memory}

Like SAT solvers, each SAT memory instance is also represented as a virtual web I/O device. The reason for such design choice is to avoid introducing a new memory type, which could complicate the web kernel architecture. Unlike web memory, SAT memory is not used in all web processes. Besides, depending on the normal form used to represent the clauses (e.g., CNF or ANF), there could be different types of SAT memory\footnote{In this paper, we cover CNF only.}. In addition, certain web methods (such as converters between normal forms) may need multiple types of SAT memory at the same time.


In this section, we focus on SAT memory with clauses given in the conjunctive normal form (CNF) since it is used in the DIMACS file format, a {\em de facto} standard supported by the majority of SAT solvers. Besides, it is used at SAT competitions\footnote{\url{https://satcompetition.github.io}}. 

SAT memory instances with CNF clauses are represented as instances of the \verb"SatCnf" web class in the root web memory. Unlike instances of \verb"SatSolver", which are preconfigured in advance, \verb"SatCnf" objects can be created and deleted dynamically. Such an approach resembles the process of attaching and detaching external data storage (e.g., USB drives) to the PC.

The CNF SAT memory consists of free variables $x_1, x_2,\dots, x_n$ and CNF clauses in the form $(l_1,l_2,\dots,l_k)$, where each $l_i, 1\le l_i\le n,$ is an integer representing a literal:
$$
\begin{array}{cc}
     x_{l_i}& \mbox{if~}l_i>0\\
     \neg x_{l_i}& \mbox{if~}l_i<0
\end{array}
$$
The number of free variables can be increased at runtime. Additional variables can be introduced, for example, to avoid exponential formula growth while converting Boolean formulas to CNF (resulting in an equisatisfiable formula, not equivalent). Besides, specific formula fragments can be easier expressible by introducing helper variables.

\verb"SatCnf" objects are created by specifying the initial number of Boolean variables.

The \verb"SatCnf" web class implements the following web methods (in a thread-safe way):
\begin{itemize}
    \item \verb"addVariable({})" - adds a new free variable and returns its index (a positive integer);
    \item \verb"addClause({clause})" - adds a new CNF clause as an array of integers $(l_1,l_2,\dots)$ representing literals, where $1\le |l_i|\le \#(\mbox{free variables})$. In order to avoid duplicate clauses, we sort literals before adding the clause. However, a Bloom filter could also be used instead.
    \item \verb"clauses({})" - returns the array of clauses currently stored in the SAT memory, e.g., \verb"[[1,-2,3],[4,-5]]";
    \item \verb"fork({detach})" - returns a new \verb"SatCnf" instance that retains the clauses of the current SAT memory with the ability to add new clauses to the forked SAT memory only.\footnote{Technically, in order to avoid copying of clauses, the common part with the original SAT memory is factored out and stored only once.}
    If \verb"detach=true", further changes to the original SAT are not reflected in the fork. Otherwise, the fork will share free variables with the origin and will get clause updates.
    
    Forks are helpful for defining SAT subtasks that can be passed to \verb"solve". For example, specific variable assignments can be added as single-literal clauses. Learned clauses (in CDCL-based solvers) can also be added to the forked memory.
\end{itemize}

Notice that we do not have methods to read and write the assignments of the variables in SAT memory. Although counter-intuitive, that protects SAT memory from possible collisions between different variable assignments by different solvers (if the solution is not unique, or in case of a partial solution).

While the methods above are convenient to constructing the initial SAT instance, invoking them via web calls involves certain serialization/deserialization overhead, which is undesirable when multiple solvers access SAT memory.
That would also negatively impact portfolio-based solvers, which need to exchange learned clauses at runtime.

In order to minimize the web kernel overhead, we introduce {\em direct access} to SAT memory, resembling the DMA\footnote{Direct memory access} feature in traditional hardware computing systems. We implement it by means of web sockets\footnote{the extension of the HTTP protocol for highly-efficient bi-directional communication via the network}.

Wes sockets of a SAT memory instance can be accessed by the address available in the \verb"directUrl" attribute.
All web socket connections are managed by an internal hub of the \verb"SatCnf" class. When a new variable or clause is added via one web socket, the hub forwards them to other web sockets. This way, new variables and clauses can be efficiently exchanged between all other nodes using the same SAT memory.

In a nutshell, the protocol used in web sockets uses binary messages corresponding to the \verb"SatCnf" methods. However, \verb"addVariable" and \verb"addClauses" can be both sent and received. When received, the client shall treat them as variable and clause synchronization messages from other solvers.

In addition to the \verb"addVariable" message, we introduce also the messages (sent by the client) for:
\begin{itemize}
    \item locking variables (thus, the last used variable index stays fixed);
    \item adding multiples free variables;
    \item unlocking variables.
\end{itemize}
These methods skip multiple round-trips when multiple free variables have to be added, and their indices used to construct clauses. Variables should be locked only for a short period of time since \verb"addVariable" calls sent by other SAT memory users will wait until variables are unlocked.

\section{Usage Example}
\label{sec:usage}
In this section, we consider the integer factorization problem and its transformation to the SAT problem as a use case. For simplicity, we consider products of exactly two integer factors ($> 1$) of the same length $l$. If both factors are primes, our transformation generates a SAT instance having the unique solution.

{\em The input:} the length $l$ and $2l$ bits of the product.

{\em The output:} A SAT instance in the SAT memory.

When the SAT memory is filled, the \verb"solve" web method of the first available portfolio-based SAT solver (found in the root web memory) is invoked to obtain the SAT assignment. The first $2l$ Boolean values of the assignment will correspond to the bits of the two factors in question.

\subsection{The Karatsuba algorithm}
In order to asymptotically minimize the number of generated variables and clauses, we apply the Karatsuba multiplication algorithm, which is based on the recursive "divide and conquer" approach \cite{knuth_vol2}.

If $u=(u_{2n_-1}\dots u_1u_0)$ and $v=(v_{2n-1}\dots v_1v_0)$ are $2n$-bit integers, they can be written as
$$
u=2^nU_1+U_0 \mbox{~and~} v=2^nV_1+V_0
$$
(the most significant bits are on the left). Then
\begin{equation}
\label{karatsuba_eq}
uv=(2^{2n}+2^n)U_1V_1+2^2(U_1-U_0)(V_0-V_1)+(2^n+1)U_0V_0.
\end{equation}
The Karatsuba algorithm is applied recursively to each of the three multiplications of $n$-bit numbers from (\ref{karatsuba_eq}). Since the algorithm expects the even number of bits at each level, we start with $u$ and $v$ represented as $2^k$-bit numbers (lacking leading zeroes are prepended).

\subsection{Initializing web memory}

In web memory, an $m$-bit integer will be represented as an ordered list of $m$ literals: for a positive literal $x_i$, the $i$-th bit value will correspond to the $x_i$ value (0 for \verb"false" and 1 for \verb"true"); for a negative literal $\neg x_i$, the $i$-th bit will correspond to the negation of $x_i$. Thus, one variable can be re-used as a positive or negative literal.

Since we are given the number of bits in each factor ($l=2^k$ by the reasons explained above), we initialize a new SAT memory instance with $2l = 2^{k+1}$ free variables. Thus, factors $u$ and $v$ are represented as the literal lists $x_{l-1},\dots x_2,x_1$ and $x_{2l-1},\dots x_{l+1},x_{l}$. 

\subsection{Generating the SAT formula}

The SAT formula generation process introduces additional integers
(additional free variables are added when needed). Those additional integers result from existing integers by applying transformation functions.

Transformation functions are constructed as Boolean expressions that bind literals of the source and target integers. Boolean expressions are built from Boolean primitives And, Or, and Not, as well as from auxiliary Boolean Xor, Majority (of 3 elements), Implication, and Equivalence, which can be easily implemented through the Boolean primitives.

Here is the list of transformation functions used as the building blocks. $L_1, L_2, \dots$ are lists of literals used to represent the corresponding source or target integers denoted as $I_1, I_2, \dots$:
\begin{itemize}
    \item $L_1$\verb".negation()"$\to L_2$: transforms $I_1$ to $I_2=-I_1$ represented as the two's complement in the $|L_1|$-bit notation. Technically, we inverse all bits of $I_1$ (by negating all literals of $L_1$) and add $1$: the $i$-th bit of $I_1$ is added to the $i$-th carry bit by means of the half adder formula (from digital logic).
    \item $L_1$\verb".sum_with"$(L_2)\to L_3$ (assume $|L_1|=|L_2|$): transforms $I_1$ and $I_2$ to $I_3=I_1+I_2$ in the $|L_1|$-bit two's complement notation. The overflow bit of $I_3$ (if any) is ignored. Technically, we use 1 half adder formula and $L_1-1$ adder formulas.
    \item $L_1$\verb".product_with"$(L_2)\to L_3$ (assume $|L_1|=|L_2|=2^k, |L_3|=2\cdot 2^k$):
    transforms two $2^k$-bit integers $I_1$ and $I_2$ (in the $2^k$-bit two's complement notation) to a $2^{k+1}$-bit integer $I_3=I_1\cdot I_2$ in the $2^{k+1}$-bit two's complement notation according to Equation~(\ref{karatsuba_eq}). Some technical nuances are:
    \begin{itemize}
        \item representing the difference of 2 numbers by means of \verb"sum_with" and \verb"negation";
        \item determining the sign of the inner Karatsuba product $(U_1-U_0)(V_0-V_1)$ from Equation~(\ref{karatsuba_eq});
        \item propagating the sign bit from the $2^k$-bit to the $2^{k+1}$-bit two's complement notation.
    \end{itemize}
\end{itemize}

By applying the transformation functions above we can construct a (non-CNF) SAT formula that binds the initial $2l$ free variables with the $2l$ literals of the product (the topmost transformation, obviously, will be \verb"product_with" on two initial lists $x_{l-1},\dots x_2,x_1$ and $x_{2l-1},\dots x_{l+1},x_{l}$). Our implementation of transformation functions introduces additional variables and re-uses literals for recurring formulas.

After constructing the formula, we convert it to CNF using the following patterns:
\begin{itemize}
    \item re-phrasing auxiliary Boolean functions (such as Xor and Majority) using And, Or, and Not;
    \item applying De Morgan's laws in order to get rid of factored-out negations;
    \item introducing new variables and equivalences for inner conjunctions. For example, given the formula $(x_1\vee (x_2\& x_3\& x_4))$, we replace $(x_2\& x_3\& x_4)$ with a new variable $x_5$, and add the equivalence $x_5\equiv x_2\& x_3\& x_4$;
    \item introducing new variables for equivalences (from the previous pattern) having more than two literals on the right side, e.g., $x_6$ to represent $x_3\& x_4$ (with the corresponding equivalence $x_6\iff x_3\&x_4$;
    \item constructing CNF clauses from 3-literal equivalences (such as \mbox{$x_5\iff x_2\& x_6$} and \mbox{$x_6\iff x_3\&x_4$}
    above) with the help of the truth table.
\end{itemize}

Afterward, we extend the CNF by appending a conjunction of literals corresponding to the product bits. The corresponding single-literal clause is added if the product bit is 1, and the literal is negated if the bit is 0.

In order to exclude the case when one the factors is 1, we add two clauses $(x_{l-1}\vee\dots x_3\vee x_2)$ and $(x_{2l-1}\vee\dots x_{l+2}\vee x_{l+1}$), meaning that, for each factor, there must be at least one bit set to 1, not counting the right-most bit ($x_1$ or $x_l$).

We exclude negative factors by adding two single-literal clauses $\neg x_{l-1}$ and $\neg x_{2l-1}$, meaning that the most-significant bit of each factor is 0 (it must be 0 for non-negative numbers in two's complement notation).

In order to specify the unique solution when the product consists of two primes, we use the condition $u-v\ge 0$. The clauses for it can be obtained in a similar manner as for the main SAT formula.

The transformation functions listed above (applied to integers represented as lists of literals) can be easily mapped to any OOP-based programming language. Access to web memory and SAT solvers (by means of web methods of the corresponding web I/O devices) is also OOP-based. Thus, SAT instances can be generated, placed into web memory, and then passed to a solver in a truly OOP way, without the need to think about the network. Furthermore, our building blocks can serve as a basis for building larger transformation functions.

The current version of the SAT generator (work-in-progress) is available at the IMCS GitHub page \url{https://github.com/LUMII-Syslab/sat-generator}.


\section{Related Work}

The invention of advanced heuristics, search space pruning, and intelligent pre-processing of SAT clauses have drastically increased the performance of SAT solvers intended to run on a single CPU. Such solvers are primarily based on the DPLL\footnote{the Davis–Putnam–Logemann–Loveland algorithm} variants CDCL\footnote{Conflict-driven clause-learning} and VSIDS\footnote{Variable State Independent Decaying Sum}.
Much research has also been done in the parallelization direction, where the three main approaches are 1) running multiple sequential SAT solvers with different settings, 2) partitioning the search space into (disjoint) subtasks and running multiple solver instances on them, and 3) the portfolio-based approach, where different types of solvers are working in parallel, with the ability to diversify them and exchange learned clauses between them \cite{hamadi_manysat_2009,balyo_hordesat_2015}.

In our approach, each SAT solver is viewed as a serial one (even if it internally uses parallelization). However, multiple solvers (of different types or the same) are also possible, and the web kernel (a part of webAppOS) allows integrating them to support different types of parallelization.

SAT memory is our innovative solution to provide distributed memory tailored to solving the SAT problem. In our implementation of the SAT memory, we use the centralized approach (as opposed to the distributed one) to deal with the coherence problem \cite{li_memory_1989}. However, our bi-directional web sockets allow web memory clients to exchange clauses, a step towards a hybrid approach. Besides, that also allows us to avoid callbacks.
The direct access to SAT memory resembles how shared character devices (from the /dev directory) are accessed in Linux and BSD systems.

Our APIs for solvers and the SAT memory resemble Portfolio Solver Interface used by HordeSAT, as well as MiniSAT's external interface \cite{balyo_hordesat_2015}. However, we separate methods related to SAT solving from the ones used to access clauses in web memory. Besides, both our APIs (for solvers and  the SAT memory) follow the pure OOP principles \cite{bib_object_thinking}. Furthermore, HordeSat uses the same \verb"setSolverInterrupt" call for both pausing and canceling the search (MiniSAT interface does not have such capability). In contrast, we use distinct methods, since in the former case, the call stack has to be maintained, while, in the latter case, we can free all the resources used by the solver. 

The integer factorization problem is one of the approaches used to generate hard SAT instances. ToughtSAT\footnote{In 2022, it is available at \url{https://github.com/joebebel/toughsat.git}; older references point to \url{https://toughsat.appspot.com/}, which is no more available.} is a collection of SAT generators, which includes a generator for SAT instances for the integer factorization problem. It has been used in SAT Competition 2019 \cite{bebel_harder_2019}. Sadly, although the generator generates succinct SAT instances, it emits errors on some factors and generates SAT instances with a non-unique solution.

A more generic approach to SAT instance generation is to use SAT compilers. Picat-SAT, FznTiny, and MiniZinc-SAT are good representatives
\cite{zhou_picat-sat_2016,bib_FznTiny,nethercote_minizinc_2007}. Although some SAT compilers use binary log encoding for integers, none of the known generic SAT compilers use the Karatsuba algorithm to represent integer products.

Our conversion to CNF utilizes the pattern of constructing CNF clauses used in the proof that SAT is reducible to 3-CNF-SAT \cite{aho_design_1974}.


\section{Conclusion}

We have proposed a software architecture where SAT solvers are shared in the distributed environment by means of the webAppOS web kernel. In our approach, any type of SAT solver can be used as a shared solver; the only requirement is to represent it as \verb"SatSolver" subclass and implement the API (web methods) from Section~\ref{sec:solver_api} (usually, glue code is sufficient for that). Our design is suitable for a single shared SAT solver, as well as for multiple SAT solvers launched with different settings or on parallel subtasks.

Since SAT solvers are represented as independent webAppOS I/O devices, they can be either on-premise (e.g., installed on specific hardware) or cloud-based. That opens opportunities to commercial SAT solvers available as a service.

In our architecture, a novel component is shared SAT memory, which can be accessed either as a webAppOS virtual I/O device or directly via web sockets. While we have considered SAT memory for storing CNF clauses, the architecture can be generalized to support clauses in the disjunctive normal form (DNF) and algebraic normal form (ANF). Having multiple types of SAT memory, our architecture allows the developers to create web methods that use multiple SAT memories simultaneously (e.g., for data conversion).

We also envisage further extensions of SAT memory. For example, belief propagation and survey propagation-based SAT solvers need to store real weights (called "warnings") associated with SAT variables and clauses \cite{bib_survey_propagation}.



\subsubsection{Acknowledgements} Research supported by the Latvian Council of Science, Project No. 2021/1-0479 "Combinatorial Optimization with Deep Neural Networks".

This preprint has not undergone peer review or any post-submission improvements or corrections. The Version of Record of this contribution is published in CCIS vol. 1598, "Digital Business and Intelligent Systems:
15th International Baltic Conference, Baltic DB\&IS 2022, Riga, Latvia, July 4–6, 2022, Proceedings", and is available online at \url{https://doi.org/10.1007/978-3-031-09850-5\_14}.

\bibliographystyle{splncs04}

\bibliography{references}

\end{document}